\documentclass[12pt]{article}
\usepackage{a4wide}
\usepackage{bbm}
\usepackage{amsmath}
\usepackage{amssymb}

\newcommand{\eV}{{\rm\,eV}}
\newcommand{\GeV}{{\rm\,GeV}}

\newcommand{\0}{\hbox{\kern2.5pt\vrule height 7.5pt\kern-2.5pt 0}}
\newcommand{\1}{\mathbbm{1}}
\newcommand{\C}{\mathbbm{C}}

\newcommand{\R}{\mathbbm{R}}
\newcommand{\Lor}{\mathop{\rm Lor}\nolimits}
\newcommand{\lOR}{\mathop{\rm lor}\nolimits}
\newcommand{\Bor}{\mathop{\rm Bor}\nolimits}
\newcommand{\bor}{\mathop{\rm bor}\nolimits}

\newcommand{\Tor}{\mathop{\rm Tor}\nolimits}
\newcommand{\Tra}{\mathop{\cal T}\nolimits}
\newcommand{\GL}{\mathop{\rm GL}\nolimits}
\newcommand{\SL}{\mathop{\rm SL}\nolimits}
\newcommand{\sL}{\mathop{\rm sl}\nolimits}
\newcommand{\SO}{\mathop{\rm SO}\nolimits}
\newcommand{\sO}{\mathop{\rm so}\nolimits}

\newcommand{\su}{\mathop{\rm su}\nolimits}

\newcommand{\tr}{\mathop{\rm tr}\nolimits}
\newcommand{\real}{\mathop{\rm Re}\nolimits}

\newcommand{\diag}{\mathop{\rm diag}\nolimits}
\newcommand{\eps}{\varepsilon}
\newcommand{\ID}{\hbox{$I\kern-3pt D$}}
\newcommand{\IID}{\hbox{$I\kern-3pt I\kern-3pt D$}}

\newcommand{\kest}{\mathop{\rm span}\nolimits}
\newcommand{\sol}{\mathop{\rm sol}\nolimits}

\newcommand{\longvec}{\raise7pt\hbox to0pt{$\,-\kern-7pt\longrightarrow$\hss}}

\newcommand{\po}{\raise7pt\hbox to0pt{$\scriptstyle\,\circ$\hss}p}
\newcommand{\Lambdao}{\raise7pt%
  \hbox to0pt{\kern1.8pt$\scriptstyle\circ$\hss}\Lambda}

\begin{document}

\thispagestyle{empty}

\begin{center}
{\Large\bf A solvable algebra for massless fermions}\\[1.0cm]
{\large Stefan Groote and Rein Saar}\\[0.3cm]
  Loodus- ja t\"appisteaduste valdkond, F\"u\"usika Instituut,\\[.2cm]
  Tartu \"Ulikool, W.~Ostwaldi 1, 59411 Tartu, Estonia
\end{center}

\vspace{0.2cm}
\begin{abstract}\noindent
We derive the stabiliser group of the four-vector, also known as Wigner's
little group, in case of massless particle states, as the maximal solvable
subgroup of the proper orthochronous Lorentz group of dimension four, known as
the Borel subgroup. In the absence of mass, particle states are disentangled
into left- and right-handed chiral states, governed by the maximal solvable
subgroups ${\rm sol}_2^\pm$ of order two. Induced Lorentz transformations are
constructed and applied to general representations of particle states.
Finally, in our conclusions it is argued how the spin-flip contribution might
be closely related to the occurrence of nonphysical spin operators.
\end{abstract}

\vspace{12pt}\noindent

\vspace{12pt}\noindent

\newpage

\section{Introduction\label{sec1}}
As neutrino oscillations are observed in experiments, it seems obvious that
all fermions carry a mass. Even though the mass spectrum reaches from small
fractions of $\eV$ for neutrinos up to $175\GeV$ for the top quark, a
hierarchy waiting still for an explanation, the fact that a fermion carries a
mass allows to go to the rest frame of the particle and to observe both
left-handed and right-handed states.

Therefore, the concept of massless fermions, moving with the speed of light,
has to be considered as an approximation. This approximation holds true if
some of the masses of fermions interacting in a perturbative calculation can
be neglected compared to other, larger fermion masses. However, while assuming
a fermion to be massless, one not only obtains an essential simplification
of the calculation but also different symmetries which are not given for
fermions with small but finite mass. As an example, the breakdown of these
symmetries can cause spin-flip effects where the result in the mass-zero limit
differs from the result for massless fermions~\cite{Lee:1964is,Jadach:1984hwn,%
Kleiss:1986ct,Jadach:1987ws,Smilga:1990uq,Contopanagos:1992fm,Falk:1993tf,%
Korner:1993dy,Groote:1996nc,Dittmaier:2002nd,Groote:2009zk}. This effect can
be understood as a discontinuity in freezing the spin of the fermion. However,
to the best of our knowledge, a deeper understanding of these effects is still
missing.

In this paper, we analyse the structure of Wigner's little group for massless
particles by adding a small but essential degree of freedom, given by the fact
that the momentum vector of a massless particle defines a projective space. In
doing so, we come to the conclusion that the stabiliser subgroup is not given
by a semisimple group as for massive particles but by a solvable group. In
Sec.~\ref{sec2} we give details on the Borel subgroup as the maximal solvable
subgroup describing the stabiliser. In Sec.~\ref{sec3} we deal with the
representation space in terms of common eigenvectors which, in a natural way,
leads to the split-off of the representation space into a left- and
right-handed part, described as Kronecker sum on Sec.~\ref{sec4}. The
two-dimensional subspaces are governed by the solvable groups $\sol_2^-$ and
$\sol_2^+$ which are expressed in terms of the Chevalley basis in
Sec.~\ref{sec5}. Finally, in Sec.~\ref{sec6} we give our conclusions and
present an outlook on how the Weyl equations for these massless states can be
combined to a Dirac equation for fermions with mass.

\subsection{Analysis of Wigner's little group}
In his paper ``Sur la dynamique de l'\'electron'' from July
1905~\cite{Poincare:1905}, Henri Poincar\'e formulates the ``Principle of
Relativity'', introduces the concepts of Lorentz transformation and Lorentz
group, postulating the covariance of the laws of nature under Lorentz
transformations. The full Lorentz group is a six-dimensional, noncompact
and non-abelian real Lie group which is not connected. The four connected
components of this group are related to each other via discrete
transformations (parity and time reversal). None of these components is simply
connected. In describing physics, one usually considers the component
connected to the identity, called the proper orthochronous Lorentz group
$\Lor(1,3)$.

An important subgroup of $\Lor(1,3)$ that preserves a given four-vector $p$
is Wigner's little group. For $p$ describing the momentum of a massive
particle, the condition $\Lambda_pp=p$ for the elements $\Lambda_p$ of the
little group can be solved in the rest frame of the particle where the
normalised momentum vector is given by $\hat p=(1;0,0,0)^T$, leading to the
block structure
\begin{equation}
\hat\Lambda_p=\begin{pmatrix}1&\vec 0^T\\\vec 0&D\end{pmatrix}=R_p,
\end{equation}
where $DD^T=\1_3=D^TD$. Therefore, the little group of a massive particle
is isomorphic to $\SO(3)$. However, for a massless particle, the momentum
vectors $p=(1;0,0,\eps)$ with $\eps=\pm 1$ for a movement along the $z$ axis
are projective vectors. Therefore, in solving the generalized equations
$\Lambda p=\lambda p$ and $\Lambda^T\eta p=\lambda^{-1}\eta p$ for
$p=(1;0,0,\eps)$ with a general value of $\eps$ and the Minkowskian metric
$\eta=\diag(1;-1,-1,-1)$ via the block ansatz
\begin{equation}
\Lambda=\begin{pmatrix}A&\vec B^T\\\vec C&D\end{pmatrix}
\end{equation}
leads to $\eps B_3=\lambda-A$, $\eps C_3=A-\lambda^{-1}$,
$\eps D_{33}=\eps\lambda-C_3$ and $\eps D_{33}=B_3+\eps\lambda^{-1}$. The two
last conditions are in agreement if and only if
\begin{equation}\label{branch}
\eps^2\lambda-A+\lambda^{-1}=\lambda-A+\eps^2\lambda^{-1}\quad
  \Leftrightarrow\quad(1-\eps^2)(\lambda-\lambda^{-1})=0.
\end{equation}
This equation marks the point where two different paths are possible to follow:
for $\lambda=\lambda^{-1}=1$ ($\lambda>0$ for the proper orthochronous Lorentz
group) one ends up again with Wigner's little group $\SO(3)$. For massless
particles, however, one has $\eps^2=1$ and, therefore, one can keep
$\lambda>0$ arbitrary, ending up with the Borel subgroup explained in the
following.

\subsection{Justification of the extension}
The introduction of an extension of Wigner's little group needs justification.
Wigner introduces the little group as a stabiliser group with respect to the
momentum vector $p$. However, because the four-length of the momentum vector
for a massless particle is zero and, therefore, the multiplication of this
vector with an arbitrary scale does not change the physics of this particle,
the physical situation is better described by a projective space. The existence
of an invariant subspace is guaranteed by the Lie--Kolchin theorem,
\subsubsection*{Lie--Kolchin theorem}
{\it If $G$ is a connected and solvable linear algebraic group defined over
an algebraically closed field and $\rho:G\to\GL(V)$ is a representation on a
nonzero finite-dimensional vector space $\,V\!$, then there exists a
one-dimensional linear subspace $L$ of $\,V\!$ such that
\begin{equation}
\rho(G)(L)=L.
\end{equation}}
In 1956, Armand Borel generalised the Lie--Kolchin theorem as a fixed-point
theorem for algebraic varieties~\cite{Borel:1956} and, therefore, also for the
projective space,
\subsubsection*{Borel fixed-point theorem}
{\it If $G$ is a connected, solvable, algebraic group acting regularly on a
non-empty, complete algebraic variety $V$ over an algebraically closed field,
then there exists a fixed-point of $\,V\!$.}

\vspace{12pt}\noindent
As expressed by Eq.~(\ref{branch}), the projectivity of the fixed-point is
broken if $\eps^2<1$, i.e.\ if the particle gains mass. In this case we are
falling back to Wigner's little group. The extension can be understood also
on the level of Lie algebras, as for massless particles the interchange of
space and time components of the momentum vector is an additional symmetry
which is absent for massive particles. Note in this context that also $E(2)$
as the little group for massless particles proposed by Wigner is a solvable
group, though not maximal.

\section{The Borel subgroup $\Bor(1,3;p)$\label{sec2}}
From now on we use $\eps$ only as the sign of the momentum 3-component. The
fact that the momentum vector $p\sim(1;0,0,\eps)$ for a massless particle is
symmetric (up to the sign $\eps=\pm1$) under the interchange of the first and
the last component gives an additional element of the algebra which is missing
so far in Wigner's little group. In order to see this, one can find solutions
for the character problem (summation over repeated indices is implied)
\begin{equation}
\Lambda^\mu{}_\nu(p)p^\nu=\lambda(\Lambda)p^\mu.
\end{equation}
Solving this problem for the Lorentz matrix
$\Lambda_p=(\Lambda^\mu{}_\nu(p))$ with $\Lambda^T_p\eta\Lambda_p=\eta$ one
obtains\footnote{The matrix is transposed (indicated by the upper index $T$)
for reasons of visualisation only.}
\begin{equation}\label{Lamexpl}
\Lambda_p=\begin{pmatrix}\cosh t+\frac12e^{-t}(u^2+v^2)&\eps u&\eps v
  &\eps(\sinh t+\frac12e^{-t}(u^2+v^2))\\
  \eps(u\cos w-v\sin w)&\cos w&-\sin w&u\cos w-v\sin w\\
  \eps(u\sin w+v\cos w)&\sin w&\cos w&u\sin w+v\cos w\\
  \eps(\sinh t-\frac12e^{-t}(u^2+v^2))&-u&-v
  &\cosh t-\frac12e^{-t}(u^2+v^2)\end{pmatrix}^T,
\end{equation}
where we have chosen $\lambda=e^t$ and introduced three additional parameters
$u$, $v$ and $w$. Expanding in these parameters one obtains
$\Lambda_p\approx\1_4+Tt+Uu+Vv+Ww$, where
\begin{eqnarray}\label{TUVW}
T=\frac{\partial\Lambda}{\partial t}\Big|_0
  =\begin{pmatrix}0&0&0&\eps\\ 0&0&0&0\\ 0&0&0&0\\ \eps&0&0&0\end{pmatrix},&&
U=\frac{\partial\Lambda}{\partial u}\Big|_0
  =\begin{pmatrix}0&\eps&0&0\\ \eps&0&0&-1\\ 0&0&0&0\\ 0&1&0&0\end{pmatrix},
  \nonumber\\[7pt]
V=\frac{\partial\Lambda}{\partial v}\Big|_0
  =\begin{pmatrix}0&0&\eps&0\\ 0&0&0&0\\ \eps&0&0&-1\\ 0&0&1&0\end{pmatrix},&&
W=\frac{\partial\Lambda}{\partial w}\Big|_0
  =\begin{pmatrix}0&0&0&0\\ 0&0&1&0\\ 0&-1&0&0\\ 0&0&0&0\end{pmatrix},\qquad
\end{eqnarray}
and the lower index ``$0$'' symbolises the initial value $t=u=v=w=0$. $T$,
$U$, $V$ and $W$ are generators of the maximal solvable Lie subgroup of
$\Lor(1,3)$, i.e.\ the Borel subgroup $\Bor(1,3;p)\subset\Lor(1,3)$. For the
corresponding Lie algebra $\mathfrak g=\kest_\R\{T,U,V,W\}=\bor(1,3;p)$ one
easily obtains
\begin{equation}
[T,U]=U,\quad [T,V]=V,\quad [W,U]=-V,\quad [W,V]=U,
\end{equation}
with all other commutators being zero. Accordingly, one has
$[\mathfrak g,\mathfrak g]=\kest_\R\{U,V\}$ and
$[[\mathfrak g,\mathfrak g],[\mathfrak g,\mathfrak g]]=0$, so that
$\mathfrak g$ is solvable. Note that the element~(\ref{Lamexpl}) of the
Borel subgroup $\Bor(1,3;p)$ is given by a polar decomposition, i.e.\ it can
be restored by calculating
\begin{equation}\label{expUVexpTW}
\Lambda_p=\exp(Uu+Vv)\exp(Tt+Ww).
\end{equation}
Because of $[T,W]=0$, one has
$\exp(Tt)\exp(Ww)=\exp(Tt+Ww)=\exp(Ww)\exp(Tt)$. The two parts of the second
exponential factor commutate with each other. They constitute the maximal
torus $\Tor(1,1;p)$ describing the transformations that leave the direction of
the momentum vector $p$ invariant: a boost directed along the $z$ axis
described by $\exp(Tt)$, and a rotation about the $z$ axis described by
$\exp(Ww)$. However, these two factors do not commute with the first
exponential factor $\Lambda_{u,v}:=\exp(Uu+Vv)$ which constitutes the
physically nontrivial part $\Tra(2;p)$ of the Borel subgroup (translations),
\begin{equation}\label{Lambdauv}
\Lambda_{u,v}=\exp\begin{pmatrix}0&\eps u&\eps v&0\\ \eps u&0&0&-u\\
  \eps v&0&0&-v\\ 0&u&v&0\end{pmatrix}
  =\begin{pmatrix}1+\frac12(u^2+v^2)&\eps u&\eps v&-\frac12\eps(u^2+v^2)\\
  \eps u&1&0&-u\\ \eps v&0&1&-v\\
  \frac12\eps(u^2+v^2)&u&v&1-\frac12(u^2+v^2)\end{pmatrix}.
\end{equation}
Note that due to the solvability, the series expansion breaks at the second
order. Together, these two parts of the polar decomposition of $\Lambda_p$
represent the Borel subgroup as a semidirect product,
\begin{equation}
\Bor(1,3;p)=\Tra(2;p)\rtimes\Tor(1,1;p).
\end{equation}

\subsection{A bridge from massive to massless}
Even though the main emphasis of this paper is layed on an independent
treatment of the little group of massless particles as the maximal noncompact
solvable subgroup of the proper orthochronous Lorentz group, there is still a
way to find a bridge connecting this part of the Lorentz group to the maximal
compact simple subgroup which is quite remarkable. Starting with a massive
particle, in the rest frame of this particle a proper orthochronous Lorentz
transformation $\hat\Lambda_p=B_{r,s,t}R_{u,v,w}$ can be written as polar
decomposition of the Wigner rotation matrix $R_{u,v,w}$ followed by a boost
$B_{r,s,t}$, where
\begin{equation}
B_{r,s,t}=\exp\begin{pmatrix}0&r&s&t\\ r&0&0&0\\
  s&0&0&0\\ t&0&0&0\end{pmatrix}, \qquad
R_{u,v,w}=\exp\begin{pmatrix}0&0&0&0\\ 0&0&w&-u\\
  0&-w&0&-v\\ 0&u&v&0\end{pmatrix}.
\end{equation}
The transformation to the laboratory frame where the momentum vector of the
particle is given by $p$ is performed with the help of the boost matrix
$B_p=B_{0,0,\eps\xi_p}$ parametrised by the momentum vector $p$,
\begin{equation}
B_p=\exp\begin{pmatrix}0&0&0&\eps\xi_p\\ 0&0&0&0\\
  0&0&0&0\\ \eps\xi_p&0&0&0\end{pmatrix}
  =\begin{pmatrix}c_p&0&0&\eps s_p\\ 0&1&0&0\\
  0&0&1&0\\ \eps s_p&0&0&c_p\end{pmatrix},
\end{equation}
where $c_p=\cosh\xi_p$ and $s_p=\sinh\xi_p$ with rapidity $\xi_p$. Accordingly,
the proper orthochronous Lorentz transformation in the laboratory frame is
given by
\begin{equation}
\Lambda_p=B_pB_{r,s,t}R_{u,v,w}B_p^{-1}=B_pB_{r,s,t}B_p^{-1}
  B_pR_{u,v,w}B_p^{-1}.
\end{equation}
For the generic Lie algebra element generating the boost $B_{r,s,t}$ one
obtains
\begin{equation}\label{BrstB}
B_p\begin{pmatrix}0&r&s&t\\ r&0&0&0\\ s&0&0&0\\ t&0&0&0\end{pmatrix}B_p^{-1}
  =\begin{pmatrix}0&c_pr&c_ps&t\\ c_pr&0&0&-\eps s_pr\\
  c_ps&0&0&-\eps s_ps\\ t&\eps s_pr&\eps s_ps&0\end{pmatrix}.
\end{equation}
Because $c_p,s_p\to\infty$ in the massless limit, $r$ and $s$ (but not
$t$) have to be renormalized in order to obtain a finite matrix
$B_pB_{r,s,t}B_p^{-1}$. This can be done by replacing $r$ by $xr$ and $s$ by
$xs$ where $xc_p=xs_p\to 1$ in the massless limit $x\to 0$. Raising the
generic element in Eq.~(\ref{BrstB}) to the exponent, one obtains
$B_{xr,xs,t}=B_{xr}B_{xs}B_t$, where the exponential factors $B_{xr}$ and
$B_{xs}$ factorise and commute with each other and with the remaining factor
$B_t$ due to the smallness of the renormalised parameters $xr$ and $xs$. The
factor $B_t$ describes a boost along the $z$ axis. Compared to the boost $B_p$
in the same direction, the former is negligible in the massless limit.
Therefore, one can replace $B_p$ with $B_pB_t^{-1}=B_t^{-1}B_p$ and obtain
\begin{equation}
\Lambda_p=B_pB_{\eps xr}B_{\eps xs}B_p^{-1}B_pR_{u,v,w}B_p^{-1}B_t\to
  \Lambda_{\eps r,\eps s}B_pR_{u,v,w}B_p^{-1}B_t.
\end{equation}
Because of the renormalisation, $B_pB_{xr}B_{xs}B_p^{-1}$ is finite in the
massless limit and gives $\Lambda_{\eps r,\eps s}$ which can be seen by
comparing the result of the exponentiation with Eq.~(\ref{Lambdauv}).

Looking at the second main factor in $\Lambda_p$, a similar consideration can
be made for $B_pR_{u,v,w}B_p^{-1}B_t$. Starting from
\begin{equation}\label{BuvwB}
B_p\begin{pmatrix}0&0&0&0\\ 0&0&w&-u\\ 0&-w&0&-v\\ 0&u&v&0\end{pmatrix}B_p^{-1}
  =\begin{pmatrix}0&\eps s_pu&\eps s_pv&0\\ \eps s_pu&0&w&-c_pu\\
  \eps s_pv&-w&0&-c_pv\\ 0&c_pu&c_pv&0\end{pmatrix},
\end{equation}
$u$ and $v$ (but not $w$) have to be renormalized using again $x$ with
$xc_p=xs_p\to 1$. Raising the generic element in Eq.~(\ref{BuvwB}) to the
exponent, one obtains $R_{xu,xv,w}=R_{xu}R_{xv}R_w$ where again all three
factors commute with each other. As $R_w$ commutes with $B_p^{-1}$ as well, this
factor can be pulled out, and the remaining product $B_pR_{xu}R_{xv}B_p^{-1}$
gives $\Lambda_{u,v}$ in the massless limit. Therefore, in this limit,
$\Lambda_p$ will decay into
\begin{equation}
\Lambda_p\to\Lambda_{\eps r,\eps s}\Lambda_{u,v}R_wB_t,\qquad
  R_w=\exp(Ww),\quad B_t=\exp(Tt).
\end{equation}
In this product, $\Lambda_{u,v}R_wB_t$ constitutes the generic element of the
Borel subgroup $\Bor(1,3;p)$ and $\Lambda_{\eps r,\eps s}$ constitutes the
rest class $\Lor(1,3)/\Bor(1,3;p)$. To conclude, the little groups of massive
and massless particles are connected by a singular transformation, induced by
an infinitesimal boost, interpreted as contraction in the sense of Inonu and
Wigner~\cite{Inonu:1953}.

\section{Common (pseudo)eigenvectors\label{sec3}}
The exponential representation~(\ref{expUVexpTW}) is a special case of the
representation
\begin{equation}\label{Lamexp}
\Lambda(\omega)=\exp\left(-\frac12\omega_{\alpha\beta}e^{\alpha\beta}\right)
\end{equation}
of the full Lorentz group where $(e^{\alpha\beta})^\mu{}_\nu
=\eta^\alpha{}_\nu\eta^{\beta\mu}-\eta^{\alpha\mu}\eta^\beta{}_\nu$,
$\eta=\diag(1,-1,-1,-1)$. Of course then, the generators $T$, $U$, $V$ and $W$
can be expressed in terms of the $e^{\alpha\beta}$,
\begin{equation}
T=-\eps e^{03},\qquad
U=e^{31}-\eps e^{01},\qquad
V=e^{32}-\eps e^{02},\qquad
W=e^{12}
\end{equation}
with the non-vanishing parameters $\eps\omega_{03}=\eps\omega_{30}=t$,
$\eps\omega_{01}=\eps\omega_{10}=\omega_{13}=-\omega_{31}=u$,
$\eps\omega_{02}=\eps\omega_{20}=\omega_{23}=-\omega_{32}=v$ and
$\omega_{21}=-\omega_{12}=w$. For technical reasons, instead of $\{T,U,V,W\}$
we may use the notation $\{T_0^\eps,T_1^\eps,T_2^\eps,T_3\}=\{T_i^\eps\}_0^3$
in the following. The upper index $\eps$ indicates the dependence on $\eps$,
where $T_0^{-\eps}=-T_0^\eps$ and $T_3^{-\eps}=T_3^\eps$. Because $T_3^\eps=W$
does not depend on $\eps$, one can skip the index in this case.

According to Lie's theorem, a solvable algebra has a single common eigenvector.
Solving the equations $T_i^\eps\ell_0=\lambda_i^{(0)}\ell_0$ ($i=0,1,2,3$),
one obtains
\begin{equation}
\ell_0=(1;0,0,\eps)^T/\sqrt2,\qquad\lambda_0^{(0)}=+1,\quad
\lambda_1^{(0)}=0,\quad
\lambda_2^{(0)}=0,\quad
\lambda_3^{(0)}=0.
\end{equation}
Not very surprisingly, the common eigenvector is just given by $p$. In order
to specify the defective matrices $\widehat T_i^\eps$ of the solvable algebra,
the equations
\begin{eqnarray}
T_i^\eps\ell_1&=&\lambda_i^{(1)}\ell_1+\gamma_{1i}^0\ell_0\nonumber\\[7pt]
T_i^\eps\ell_2&=&\lambda_i^{(2)}\ell_2+\gamma_{2i}^1\ell_1+\gamma_{2i}^0\ell_0
  \nonumber\\[7pt]
T_i^\eps\ell_3&=&\lambda_i^{(3)}\ell_3+\gamma_{3i}^2\ell_2+\gamma_{2i}^1\ell_1
  +\gamma_{2i}^0\ell_0
\end{eqnarray}
are solved step-wise to obtain a system of pseudo-eigenvectors and
-eigenvalues. Collecting all these equations in a single one, after some
normalisation one obtains
\begin{equation}\label{matP}
T_i^\eps P=P\begin{pmatrix}
  \lambda_i^{(0)}&\gamma_{1i}^0&\gamma_{2i}^0&\gamma_{3i}^0\\
  0&\lambda_i^{(1)}&\gamma_{2i}^1&\gamma_{3i}^1\\
  0&0&\lambda_i^{(2)}&\gamma_{3i}^2\\ 0&0&0&\lambda_i^{(3)}\end{pmatrix},\qquad
P=\frac1{\sqrt2}\begin{pmatrix}1&0&0&-\eps\\ 0&1&i&0\\
  0&i&1&0\\ \eps&0&0&1\end{pmatrix},
\end{equation}
where $P=(\ell_0,\ell_1,\ell_2,\ell_3)$ is rearranged in order to be unitary,
$P^{-1}=P^\dagger$. Turning back to the original notation, one obtains
$TP=P\widehat T$, $UP=P\widehat U$, $VP=P\widehat V$ and $WP=P\widehat W$,
where
\begin{eqnarray}
\widehat T=\begin{pmatrix}1&0&0&0\\ 0&0&0&0\\ 0&0&0&0\\ 0&0&0&-1\end{pmatrix},&&
\widehat U=\begin{pmatrix}0&\eps&i\eps&0\\ 0&0&0&-1\\
  0&0&0&i\\ 0&0&0&0\end{pmatrix},\nonumber\\[7pt]
\widehat V=\begin{pmatrix}0&i\eps&\eps&0\\ 0&0&0&i\\
  0&0&0&-1\\ 0&0&0&0\end{pmatrix},&&
\widehat W=\begin{pmatrix}0&0&0&0\\ 0&i&0&0\\ 0&0&-i&0\\ 0&0&0&0\end{pmatrix}
\end{eqnarray}
are upper triangular forms of the four generators.

\subsection{Generating the (pseudo)eigenvectors}
Even though the four generators have only a single common eigenvector, this is
not the case for the generic element $\Lambda_p\in\Bor(1,3;p)$ in
Eq.~(\ref{Lamexpl}). Solving the fourth-order equation
$\det(\Lambda_p-\lambda\1)=0$ for $\lambda$ leads to
$\lambda\in\{e^t,e^{iw},e^{-iw},e^{-t}\}$. The corresponding system of
eigenvectors can be calculated. However, here we give a more elegant method to
calculate this system of eigenvectors. Using the exponential
representation~(\ref{expUVexpTW}) and
\begin{equation}
(Uu+Vv)P=P(\widehat Uu+\widehat Vv),\qquad
(Tt+Ww)P=P(\widehat Tt+\widehat Ww),
\end{equation}
one obtains $\Lambda_p=PK_{u,v}K_{t,w}P^{-1}$ with
unipotent $K_{u,v}=\exp(\widehat Uu+\widehat Vv)$ and
semisimple $K_{t,w}=\exp(\widehat Tt+\widehat Ww)$,
\begin{equation}
K_{u,v}=\begin{pmatrix}1&\eps(u+iv)&\eps(v+iu)&-\eps(u^2+v^2)\\
  0&1&0&-u+iv\\ 0&0&1&-v+iu\\ 0&0&0&1\end{pmatrix},\quad
K_{t,w}=\begin{pmatrix}e^t&0&0&0\\ 0&e^{iw}&0&0\\
  0&0&e^{-iw}&0\\ 0&0&0&e^{-t}\end{pmatrix}.
\end{equation}
Because $K_{t,w}$ is a diagonal matrix containing the four eigenvectors, the
system of eigenvectors is given by the matrix $Q$ obeying
$\Lambda_pQ=QK_{t,w}$. Inserting $\Lambda_p=PK_{u,v}K_{t,w}P^{-1}$ into this
eigenvalue equation, after some rearrangements one obtains
\begin{equation}
P^{-1}Q=K_{u,v}K_{t,w}P^{-1}QK_{t,w}^{-1}.
\end{equation}
This equation for the unknown quantity $P^{-1}Q$ can be solved iteratively,
starting with $P^{-1}Q=\1$, i.e.\ $Q=P$. The iterative solution can be shown
to converge to
\begin{equation}
P^{-1}Q=\begin{pmatrix}1&\displaystyle\frac{\eps(u+iv)}{1-e^{t-iw}}
  &\displaystyle\frac{\eps(v+iu)}{1-e^{t+iw}}&\displaystyle
  \frac{-\eps(u^2+v^2)}{(1-e^{t-iw})(1-e^{t+iw})}\\\noalign{\smallskip}
  0&1&0&\displaystyle\frac{-u+iv}{1-e^{t+iw}}\\\noalign{\smallskip}
  0&0&1&\displaystyle\frac{-v+iu}{1-e^{t-iw}}\\\noalign{\smallskip}
  0&0&0&1\end{pmatrix}.
\end{equation}
Multiplying with $P$ from the left, one finally obtains the system of
eigenvectors
\begin{equation}\label{Qmat}
Q=\frac1{\sqrt2}\begin{pmatrix}1&\displaystyle\frac{\eps(u+iv)}{1-e^{t-iw}}
  &\displaystyle\frac{\eps(v+iu)}{1-e^{t+iw}}&\displaystyle
  -\eps-\frac{\eps(u^2+v^2)}{(1-e^{t-iw})(1-e^{t+iw})}\\\noalign{\smallskip}
  0&1&i&\displaystyle\frac{-u+iv}{1-e^{t+iw}}
  +i\displaystyle\frac{-v+iu}{1-e^{t-iw}}\\\noalign{\smallskip}
  0&i&1&i\displaystyle\frac{-u+iv}{1-e^{t+iw}}
  +\displaystyle\frac{-v+iu}{1-e^{t-iw}}\\\noalign{\smallskip}
  \eps&\displaystyle\frac{u+iv}{1-e^{t-iw}}
  &\displaystyle\frac{v+iu}{1-e^{t+iw}}&\displaystyle
  1-\frac{\eps(u^2+v^2)}{(1-e^{t-iw})(1-e^{t+iw})}\end{pmatrix}.
\end{equation}
Expressed in a slightly philosophically manner, one can say that starting from
the very sparse boundary of four defective matrices, the Lie algebra (in this
case, the Borel subalgebra) knits the sweater $Q$ for the Lie group in a
straightforward, iterative way.

\section{Kronecker sum of solvable algebras\label{sec4}}
Although we were able to analyse the solvable algebra $\bor(1,3;p)$, the
representation in terms of the generators $T$, $U$, $V$ and $W$ is not the
best one to see the structure of this algebra. Therefore, we use a second one,
namely
\begin{eqnarray}\label{defJK}
J^\eps_3=\frac12(-T-iW),&& J^\eps_-=\frac12(-U+iV),\nonumber\\[7pt]
K^\eps_3=\frac12(T-iW),&& K^\eps_+=\frac12(U+iV),
\end{eqnarray}
obeying
\begin{eqnarray}\label{comJK}
[J^\eps_3,J^\eps_3]=0,&&[J^\eps_3,J^\eps_-]=-J^\eps_-,\qquad
  [J^\eps_-,J^\eps_-]=0,\nonumber\\[7pt]
[K^\eps_3,K^\eps_3]=0,&&[K^\eps_3,K^\eps_+]=K^\eps_+,\qquad
  [K^\eps_+,K^\eps_+]=0
\end{eqnarray}
and $[J^\eps_3,K^\eps_3]=[J^\eps_3,K^\eps_+]=[J^\eps_-,K^\eps_3]=
[J^\eps_-,K^\eps_+]=0$.  The first justification for the sign notations for
$J^\eps_-$ and $K^\eps_+$ is given by the commutator relations~(\ref{comJK}).
In terms of the pairs $\{J^\eps_3,J^\eps_-\}$ and $\{K^\eps_3,K^\eps_+\}$ of
generators $\bor(1,3;p)$ can be rewritten as a Kronecker sum
$\sol_2^-\boxplus\sol_2^+$ of two two-dimensional solvable algebras, as will
be detailed in the following.

\subsection{Weyl's unitary trick}
A deeper look at this change of representation unveils that this change is
actually a composition of several steps. In order to illustrate these steps,
one can start again with the proper orthochronous Lorentz group
$\Lor(1,3)\subset\SO(1,3)$, the elements of which are given by the exponential
representation~(\ref{Lamexp}) where $(e^{\alpha\beta})^\mu{}_\nu
=\eta^\alpha{}_\nu\eta^{\beta\mu}-\eta^{\alpha\mu}\eta^\beta{}_\nu$,
$\eta=\diag(1,-1,-1,-1)$. This representation can be written in a different
form as
\begin{equation}
\Lambda_p=\exp(\vec\tau\cdot\vec E+\vec\omega\cdot\vec B),
\end{equation}
using an analogy to the electromagnetic field strength tensor $F^{\mu\nu}$
to write
\begin{equation}
-\frac12\omega_{\alpha\beta}e^{\alpha\beta}
  =\begin{pmatrix}0&\omega_{01}&\omega_{02}&\omega_{03}\\
  \omega_{01}&0&\omega_{12}&-\omega_{31}\\
  \omega_{02}&-\omega_{12}&0&\omega_{23}\\
  \omega_{03}&\omega_{31}&-\omega_{23}&0\end{pmatrix}
  =\vec\tau\cdot\vec E+\vec\omega\cdot\vec B,
\end{equation}
where $\vec\tau=(\omega_{01},\omega_{02},\omega_{03})$,
$\vec\omega=(\omega_{23},\omega_{31},\omega_{12})$ and
$\vec E=-(e^{01},e^{02},e^{03})$, $\vec B=-(e^{23},e^{31},e^{12})$ or
$e^{0i}=-E_i$, $e^{ij}=-\epsilon_{ijk}B_k$ with the convention of lower indices
for $\vec E$ and $\vec B$ and related three-vectors, $\epsilon_{123}=1$. The
$3+3$ generators of $\Lor(1,3)$ obey the commutation relations
\begin{equation}\label{comBE}
[B_i,B_j]=\epsilon_{ijk}B_k,\qquad
[B_i,E_j]=\epsilon_{ijk}E_k,\qquad
[E_i,E_j]=-\epsilon_{ijk}B_k.
\end{equation}
Obviously, the algebra $\lOR(1,3)$ is a real algebra. It contains a compact
subalgebra $\mathfrak k$ related to the $B^i$ which is isomorphic to the
compact algebra $\sO(3)$. Actually, $\lOR(1,3)$ is in the shape of a Cartan
decomposition $\mathfrak g=\mathfrak k\ \dot+\ \mathfrak p$ characterized by
the values $\phi(\mathfrak k)=\mathfrak k$, $\phi(\mathfrak p)=-\mathfrak p$
of an involution $\phi$. As vector spaces, $\mathfrak k$ and $\mathfrak p$ are
orthogonal, because given a scalar product $(\mathfrak k,\mathfrak p)$
invariant under this involution, one obtains
\begin{equation}
(\mathfrak k,\mathfrak p)=(\phi(\mathfrak k),\phi(\mathfrak p))
  =(\mathfrak k,-\mathfrak p)=-(\mathfrak k,\mathfrak p)
  \quad\Rightarrow\quad(\mathfrak k,\mathfrak p)=0
\end{equation}
However, $[\mathfrak k,\mathfrak p]\ne 0$. Therefore, we used the symbol
$\dot+$ instead of the symbol $\oplus$ for the direct sum. The algebra
$\mathfrak g$ can be transformed to a compact form by using Weyl's unitary
trick. The result is an algebra
$\mathfrak g^*=\mathfrak k\ \dot+\ \mathfrak i\mathfrak p$, where the
implications for introducing an imaginary factor will be explained later.
In case of $\lOR(1,3)$, the involution is given by
\begin{equation}
\phi:e^{\mu\nu}\to\eta(e^{\mu\nu}):=\eta e^{\mu\nu}\eta=-e^{\mu\nu T}
\end{equation}
(matrix indices are suppressed) or $\phi(\vec B)=\vec B$,
$\phi(\vec E)=-\vec E$. Therefore, the compact form of $\lOR(1,3)$ is
given by the generators $B^i$ and $iE^i$ obeying the commutation relations
\begin{equation}\label{comBiE}
[B_i,B_j]=\epsilon_{ijk}B_k,\qquad
[B_i,(iE_j)]=\epsilon_{ijk}(iE_k),\qquad
[(iE_i),(iE_j)]=\epsilon_{ijk}B_k.
\end{equation}
Considered as a real algebra, this algebra is isomorphic to $\sO(4)$. However,
the generators are antihermitean, $B_i^\dagger=-B_i$ and $(iE_i)^\dagger=-iE_i$
and, therefore, the group is unitary. In general, Weyl's unitary trick can be
seen to lead always to unitary Lie groups.

\subsection{Duplication and complexification}
The addition of an imaginary factor $i$ turns the real algebra into a complex
algebra, at least for intermediate steps. In general, this process is called
complexification and is denoted by a lower index $\C$ (or additional argument)
to the algebra symbol. Given a real Lie algebra $L$, the {\em duplication\/}
of this algebra is given by~\cite{BauerleKerf}
\begin{equation}
L+iL:=\{x+iy\ |\ x,y\in L\}.
\end{equation}
$L+iL$ is still a real vector space. In defining the multiplication of an
element $x+iy\in L_C$ with a complex number $\alpha=a+ib\in\C$ by
$(a+ib)(x+iy):=(ax-by)+i(bx+ay)$, and the commutator of two elements
$x+iy$ and $x'+iy'$ by
\begin{equation}
[x+iy,x'+iy']:=[x,x']-[y,y']+i([x,y']+[y,x']),
\end{equation}
$L+iL$ constitutes a {\em complex form\/}, denoted by $L_\C:=\C\otimes_\R L$.
This complex form again is a complex Lie algebra which is called the
{\em complexification\/} of $L$. Applied to the actual case, the
complexification turns the real algebra $\sO(1,3)$ into the complex algebra
$\sO(4,\C)$.

However, it is obvious that the algebra given by the commutator
relations~(\ref{comBiE}) is real, not complex. The final algebra, therefore,
is a {\em real form\/} of this complex algebra, defined as follows: a
subalgebra $K$ of the duplicated algebra $L+iL$ is called {\em real form\/}
if the complexification of this subalgebra is the same as the original
algebra, $K_\C=L$. As the duplication is not unique,\footnote{For instance, a
part of the basis elements can be duplicated with $i$, another part with
$-i$.} there are also different real forms to a given complex algebra.

\subsection{Compactified and decompactified real forms}
Most important real forms are the {\em normal real form\/} where the
duplicates are again taken as separate elements, and the {\em compact real
form\/} which exists for all (semi)simple complex Lie algebras. Because we
will meet these forms in the lower-dimensional case, we postpone the
discussion about the different real forms. In the actual case, one of the
compact real forms is $\sO(4)$. However, another one is given by
\begin{equation}
A_i=\frac12(B_i+iE_i),\qquad \bar A_i=\frac12(B_i-iE_i)
\end{equation}
with the commutation rules
\begin{equation}
[A_i,A_j]=\epsilon_{ijk}A_k,\qquad
[A_i,\bar A_j]=0,\qquad
[\bar A_i,\bar A_j]=\epsilon_{ijk}\bar A_k.
\end{equation}
Therefore, the algebra decomposes into two separate algebras which are
isomorphic to $\su(2)$.\footnote{$\su(2)$ is preferred instead of $\sO(3)$
because $A_i$ and $\bar A_i$ are antihermitean, leading to unitary groups.}
Turning back to solvable groups, the decomposition into
$\su(2)=\kest_\R\{A_i\}$ and $\su(2)=\kest_\R\{\bar A_i\}$ is not yet conform
with the definitions given in Eq.~(\ref{defJK}). Looking at the definitions of
$T_i^\eps$ on the one hand and the definitions of $E_i$ and $B_i$ on the other
hand, one obtains
\begin{equation}
J^\eps_3=\frac12(\eps e^{03}-ie^{12})=\frac12(-\eps E_3+iB_3),\quad
K^\eps_3=\frac12(-\eps e^{03}-ie^{12})=\frac12(\eps E_3+iB_3)
\end{equation}
and generally $J^\eps_i=\frac12(-\eps E_i+iB_i)=i\bar A_i$,
$K^\eps_i=\frac12(\eps E_i+iB_i)=iA_i$. As $A_i$ and $\bar A_i$ are
antihermitean, $J^\eps_i$ and $K^\eps_i$ are hermitean and, therefore,
constitute decompactified subgroups generated by $\exp(ij_iJ^\eps_i$) and
$\exp(ik_iK^\eps_i)$. One obtains
\begin{eqnarray}
J^\eps_1-iJ^\eps_2&=&\frac12\left(-\eps E_1+B_2+i(\eps E_2+B_1)\right)
  \ =\ \frac12(-T_1^\eps+iT_2^\eps)\ =\ J^\eps_-,\nonumber\\
K^\eps_1+iK^\eps_2&=&\frac12\left(\eps E_1-B_2+i(\eps E_2+B_1)\right)
  \ =\ \frac12(T_1^\eps+iT_2^\eps)\ =\ K^\eps_+
\end{eqnarray}
which is the other justification for the sign notations in $J^\eps_-$ and
$K^\eps_+$. Actually, the algebra looks like $\sL(2,\R)$ with one generator
missing ($J^\eps_+$ or $K^\eps_-$, respectively). In $\lOR(1,3)$, these
missing generators exist. In $\bor(1,3;p)$, however, the generators are found
in the resp.\ other algebra with opposite sign $\eps$,
\begin{eqnarray}
J^\eps_+&=&J^\eps_1+iJ^\eps_2
  \ =\ \frac12\left(-\eps E_1-B_2+i(-\eps E_2+B_1)\right)
  \ =\ \frac12\left(T^{-\eps}_1+iT^{-\eps}_2\right)\ =\ K^{-\eps}_+,\nonumber\\
K^\eps_-&=&K^\eps_1-iK^\eps_2
  \ =\ \frac12\left(\eps E_1+B_2+i(-\eps E_2+B_1)\right)
  \ =\ \frac12\left(-T^{-\eps}_1+iT^{-\eps}_2\right)\ =\ J^{-\eps}_-,\qquad
\end{eqnarray}
while $J^{\pm\eps}_3=K^{\mp\eps}_3$. Therefore, $\bor(1,3;p)$ splits up into
the subalgebras
\begin{eqnarray}
\sol_2^-&:=&\kest_\R\{J^\eps_3,J^\eps_-\}
  \ =\ \kest_\R\{K^{-\eps}_3,K^{-\eps}_-\}
\quad\mbox{and}\nonumber\\
\sol_2^+&:=&\kest_\R\{K^\eps_3,K^\eps_+\}
  \ =\ \kest_\R\{J^{-\eps}_3,J^{-\eps}_+\}.
\end{eqnarray}
As there is a homomorphism between the two algebras $\sol_2^+$ and $\sol_2^-$,
both solvable algebras are maximal and, therefore, are Borel subalgebras of
the larger algebra $\sL(2,\R)$. For a free choice of $\eps$ one can represent
the two Borel subalgebras as being generated by a solvable part of the set
$\{J^{\pm\eps}_3,J^{\pm\eps}_+,J^{\pm\eps}_-\}$ of generators of $\sL(2,\R)$,
thereby skipping the second (redundant) set
$\{K^{\mp\eps}_3,K^{\mp\eps}_+,K^{\mp\eps}_-\}$. Alternatively, one can use
the two sets and skip $\eps=+1$. Though the first choice is more intriguing,
for this paper we stay with the clearer second one.

\subsection{In quest of left and right\label{Sec:LR}}
Searching for eigenvectors of the set $\{J_3,J_+,J_-\}$ one
finds that these eigenvectors are disjoint, as known for semisimple algebras.
The same holds for the set $\{K_3,K_+,K_-\}$. However, for each
of the solvable subalgebras $\sol_2^-$ and $\sol_2^+$ one obtains only a
single common eigenvector. In order to analyse the eigenvector structure, we
return to the eigenvectors
\begin{equation}
\ell_0=\frac1{\sqrt2}\begin{pmatrix}1\\ 0\\ 0\\ \eps\end{pmatrix},\quad
\ell_1=\frac1{\sqrt2}\begin{pmatrix}0\\ 1\\ i\\ 0\end{pmatrix},\quad
\ell_2=\frac1{\sqrt2}\begin{pmatrix}0\\ 1\\ -i\\ 0\end{pmatrix}\ \mbox{and}\
\ell_3=\frac1{\sqrt2}\begin{pmatrix}1\\ 0\\ 0\\ -\eps\end{pmatrix}
\end{equation}
of Sec.~\ref{sec3} to which we apply the algebra elements, obtaining
\begin{equation}
\begin{matrix}
J_3\ell_0=-\frac12\ell_0&J_+\ell_0=-\eps\ell_1&J_-\ell_0=0&
K_3\ell_0=+\frac12\ell_0&K_+\ell_0=0&K_-\ell_0=\eps\ell_1\\
J_3\ell_1=+\frac12\ell_1&J_+\ell_1=0&J_-\ell_1=-\eps\ell_0&
K_3\ell_1=+\frac12\ell_1&K_+\ell_1=0&K_-\ell_1=\eps\ell_3\\
J_3\ell_2=-\frac12\ell_2&J_+\ell_2=-\eps\ell_3&J_-\ell_2=0&
K_3\ell_2=-\frac12\ell_2&K_+\ell_2=\eps\ell_0&K_-\ell_2=0\\
J_3\ell_3=+\frac12\ell_3&J_+\ell_3=0&J_-\ell_3=-\eps\ell_2&
K_3\ell_3=-\frac12\ell_3&K_+\ell_3=\eps\ell_1&K_-\ell_3=0
\end{matrix}
\end{equation}
For $\{J_3,J_-\}$ the common eigenvector is given as a linear
combination of $\ell_0$ and $\ell_2$ while for $\{K_3,K_+\}$ the
common eigenvector is given by the linear combination of $\ell_0$ and $\ell_1$.
On the other hand, the common eigenvector for $\{J_3,J_+\}$ is a
linear combination of $\ell_3$ and $\ell_1$ while the common eigenvector of
$\{K_3,K_-\}$ is a linear combination of $\ell_3$ and $\ell_2$.
While $\ell_0$ ($\ell_3$) is proportional to the (space inverted) momentum
four-vector $p$, the interpretation of the eigenvectors $\ell_1$ and $\ell_2$
deserves more effort. For this one can take refuge to the circular
polarisation~\cite{Jackson,Ida}. The representation
\begin{eqnarray}
\vec E(z,t)=E_0\real\left((\vec e_x+i\vec e_y)e^{ikz-i\omega t}\right)
  =E_0\left(\vec e_x\cos(kz-\omega t)-\vec e_y\sin(kz-\omega t)\right)
\end{eqnarray}
describes the right turn of the electric vector in the $(x,y)$ plane, as can
be seen by comparing the solution for $z=0$ at $t=0$ and after a short time
$t=\Delta t$. Therefore, the vector $\ell_1$ can be identified with the right
turn. However, a turn can be identified with handedness or chirality only in
combination with a direction of propagation,\footnote{In optics this solution
is named left-polarised, as looked at in the direction the light comes from
(passive direction). In our case, however, we consider the direction of
propagation (active direction).} as in case of the circular polarisation by
the argument $kz-\omega t$. This direction is given by $\ell_0$ (or $\ell_3$).
Therefore, one can interpret (in case of $\eps=1$)
\begin{eqnarray}
\{J_3,J_-\}&&\mbox{as forward-propagating left-handed,}\nonumber\\[7pt]
\{K_3,K_+\}&&\mbox{as forward-propagating right-handed,}\nonumber\\[7pt]
\{J_3,J_+\}&&\mbox{as backward-propagating left-handed, and}\nonumber\\[7pt]
\{K_3,K_-\}&&\mbox{as backward-propagating right-handed.}
\end{eqnarray}

\subsection{The irreducible representation}
In terms of $4\times 4$ matrices the generators $J_i$ and $K_i$ ($i=3,\pm$)
are, of course, not given in the irreducible representation. However, they can
be related to irreducible representations in an easy way. In fact, there is a
similarity transformation such that
\begin{eqnarray}\label{simtra}
J_i\to S^{-1}J_iS=:J^\boxplus_i,\qquad
K_i\to S^{-1}K_iS=:K^\boxplus_i
\end{eqnarray}
(a deeper understanding of the representation index $\boxplus$ will be given
soon), where
\begin{equation}\label{matrixS}
S=\frac1{\sqrt2}
  \begin{pmatrix}0&-1&1&0\\ -1&0&0&1\\ -i&0&0&-i\\ 0&1&1&0\end{pmatrix}
\end{equation}
and $S^{-1}=S^\dagger$. In detail, one obtains
\begin{eqnarray}
J^\boxplus_3=\frac12(\sigma_3\otimes\1),&&
J^\boxplus_\pm=\frac12(\sigma_\pm\otimes\1),\nonumber\\
K^\boxplus_3=\frac12(\1\otimes\sigma_3),&&
K^\boxplus_\pm=\frac12(\1\otimes\sigma_\pm),\qquad
\end{eqnarray}
where the outer product is defined by $(A\otimes B)_{(ik)(jl)}:=A_{ij}B_{kl}$,
i.e.\ the first matrix sets the frame for the second one. The matrices
\begin{equation}
\sigma_1=\begin{pmatrix}0&1\\ 1&0\end{pmatrix},\qquad
\sigma_2=\begin{pmatrix}0&-i\\ i&0\end{pmatrix},\qquad
\sigma_3=\begin{pmatrix}1&0\\ 0&-1\end{pmatrix}
\end{equation}
are the usual Pauli matrices, and $\sigma_\pm=\sigma_1\pm i\sigma_2$.
The same similarity transformation via $S$ can be applied also to the
generators $E_i$ and $B_i$ of the proper orthochronous Lorentz group
$\Lor(1,3)$. One obtains
\begin{eqnarray}
E^\boxplus_i&:=&S^{-1}E_iS\ =\ -\frac12(\sigma_i\otimes\1-\1\otimes\sigma_i),
  \nonumber\\
B^\boxplus_i&:=&S^{-1}B_iS\ =\ -\frac i2(\sigma_i\otimes\1+\1\otimes\sigma_i).
\end{eqnarray}
These two results can be rewritten by employing the {\em Kronecker sum}
\begin{equation}
A\boxplus B:=A\otimes\1+\1\otimes B.
\end{equation}
Using this notation, one obtains
\begin{equation}
E^\boxplus_i=-\frac12\sigma_i\boxplus\left(+\frac12\sigma_i\right),\qquad
B^\boxplus_i=-\frac i2\sigma_i\boxplus\left(-\frac i2\sigma_i\right).
\end{equation}
Therefore, the representation index $\boxplus$ indicates that in this
representation obtained via the similarity transformation with $S$ the matrix
can be written as a Kronecker sum. It is characteristic that
\begin{equation}\label{JKbox}
J^\boxplus_i=\frac12\sigma_i\boxplus\0,\qquad
K^\boxplus_i=\0\boxplus\frac12\sigma_i
\end{equation}
contribute only to the first or second component of the Kronecker sum,
respectively. Following the argumentation of Sec.~\ref{Sec:LR} one can
conclude that the first component of the Kronecker sum (and thereby
$J^\boxplus_i$) is left-handed while the second component of the Kronecker sum
(and thereby $K^\boxplus_i$) is right-handed. Finally, we conclude that via
the same similarity transformation $S$ the maximal solvable algebra
$\bor(1,3;p)$ in the representation of this section can indeed be decomposed
into the Kronecker sum $\sol_2^-\boxplus\sol_2^+$.

\section{The Chevalley basis\label{sec5}}
From Eqs.~(\ref{JKbox}) it is obvious that $\sol_2^-$ and $\sol_2^+$ are
isomorphic to Borel subalgebras of the real algebra $\sL(2,\R)$, given in
the {\em Chevalley basis\/} by the three generators
\begin{equation}
\sigma_3=\begin{pmatrix}1&0\\ 0&-1\end{pmatrix},\qquad
\sigma_+=\begin{pmatrix}0&2\\ 0&0\end{pmatrix}\quad\mbox{and}\quad
\sigma_-=\begin{pmatrix}0&0\\ 2&0\end{pmatrix}.
\end{equation}
One can write $\sol_2^\pm=\kest_\R\{\sigma_3,\sigma_\pm\}$. The algebra
$\sL(2,\R)=\kest_\R\{\sigma_3,\sigma_+,\sigma_-\}$ can be complexified to
obtain $\sL(2,\C)=\kest_\C\{\sigma_3,\sigma_+,\sigma_-\}$. Therefore,
$\sL(2,\R)$ is a real form of $\sL(2,\C)$. The compact real form of
$\sL(2,\C)$ is given by $\su(2)=\kest_R\{\sigma_1,\sigma_2,\sigma_3\}$, while
$\sL(2,\R)$ can be called the decompactified real form of $\sL(2,\C)$. In a
similar way as the complexified version of $\lOR(1,3)$ is isomorphic to
$\su(2)\boxplus\su(2)$, the complexified version of the extended little
algebra $\bor(1,3;p)$ is isomorphic to $\sol_2^-\boxplus\sol_2^+$.

\subsection{Common eigenvectors}
The concept of common eigenvectors introduced in Sec.~\ref{Sec:LR} pulls
through to the very core, i.e.\ to the irreducible representation. The
set of generators $\{\sigma_3,\sigma_+\}$ of $\sol_2^+$ have the common
eigenvector $(1,0)^T$ and the set $\{1,0\}$ of eigenvalues while for
$\{\sigma_3,\sigma_-\}$ (i.e.\ $\sol_2^-$) the common eigenvector is $(0,1)^T$
with eigenvalues $\{-1,0\}$. Reintroducing the sign $\eps$, the two
non-trivial eigenvalue equations can be cast into the form
\begin{equation}
\sigma_3\psi_+=\eps\psi_+,\qquad
\psi_+=\begin{pmatrix}(1+\eps)\psi^1\\ (1-\eps)\psi^2\end{pmatrix}.
\end{equation}
This is the first quantisation step. Indeed,
introducing
\begin{equation}
\sigma_0:=\begin{pmatrix}1&0\\ 0&1\end{pmatrix}=\1
\end{equation}
one obtains the Weyl equation
($(\sigma^\mu):=(\sigma_0;\sigma_1,\sigma_2,\sigma_3)$)
\begin{equation}
0=(\eps\sigma_0-\sigma_3)\psi_+=\eps p_\mu\sigma^\mu\psi_+
  =:\eps\sigma(p)\psi_+.
\end{equation}
However, this is not the only possible quantisation. Equivalently, one may
write
\begin{equation}
\sigma_3\psi_-=-\eps\psi_-,\qquad
  \psi_-=\begin{pmatrix}(1-\eps)\psi^1\\(1+\eps)\psi^2\end{pmatrix}
\end{equation}
or
\begin{equation}
0=(\eps\sigma_0+\sigma_3)\psi_-=\eps p_\mu\tilde\sigma^\mu\psi_-
  =:\eps\tilde\sigma(p)\psi_-
\end{equation}
($(\tilde\sigma^\mu)=(\sigma_0;-\sigma_1,-\sigma_2,-\sigma_3)$) which is
the dual Weyl equation. In using the tilde notation for $\tilde\sigma$ one
avoids the breakdown of the covariant notation. Using Weyl's representation
\begin{equation}
\gamma_W^\mu=\begin{pmatrix}0&\sigma^\mu\\ \tilde\sigma^\mu&0\end{pmatrix},
  \qquad\gamma_W^5=\begin{pmatrix}-\1&0\\ 0&\1\end{pmatrix}
\end{equation}
of the Dirac matrices, for finite mass $m$ one ends up with the Dirac
equation
\begin{equation}
(p_\mu\gamma_W^\mu-mc)\psi_W=0,\qquad
\psi_W=\begin{pmatrix}\psi_-\\ \psi_+\end{pmatrix}.
\end{equation}
$\psi_+$ is the right-handed spinor and $\psi_-$ is the left-handed spinor.
This is in agreement with the usual definition
$\psi_R=\frac12(1+\gamma^5)\psi_W=(0,\psi_+)^T$ and
$\psi_L=\frac12(1-\gamma^5)\psi_W=(\psi_-,0)^T$.

\subsection{Induced Lorentz transformations}
The contractions of the momentum four-vector $p$ with $\sigma$ and
$\tilde\sigma$ induces two (proper orthochronous) Lorentz transformations
$A_\Lambda$ and $\tilde A_\Lambda$ which make the diagram
\begin{center}\begin{tabular}{rlcr}
$A_\Lambda^{\phantom\dagger}:$&$\sigma(p)$&$\longrightarrow$&
  $\sigma(\Lambda p)$\\
$\llap{$\pi$}\uparrow$&$\uparrow\rlap{$\sigma$}$&&$\uparrow\rlap{$\sigma$}$\\
$\Lambda:$&$p$&$\longrightarrow$&$\Lambda p$\\
$\llap{$\tilde\pi$}\downarrow$&$\downarrow\rlap{$\tilde\sigma$}$&&
  $\downarrow\rlap{$\tilde\sigma$}$\\
$\tilde A_\Lambda^{\phantom\dagger}:$&$\tilde\sigma(p)$&$\longrightarrow$&
  $\tilde\sigma(\Lambda p)$\\
\end{tabular}\end{center}
commutative. The induced Lorentz transformations are defined by
\begin{equation}\label{sigtrans}
A_\Lambda^{\phantom\dagger}\sigma(p)A_\Lambda^\dagger=\sigma(\Lambda p),\qquad
\tilde A_\Lambda^{\phantom\dagger}\tilde\sigma(p)\tilde A_\Lambda^\dagger
  =\tilde\sigma(\Lambda p).
\end{equation}
A long but straightforward calculation shows that
\begin{equation}\label{ALamdef}
A_\Lambda=\frac{\sigma^\mu\Lambda_{\mu\nu}\tilde\sigma^\nu}{2
  \tr(A_\Lambda^\dagger)},\quad
A_\Lambda^{-1}=\frac{\sigma^\mu\Lambda_{\nu\mu}\tilde\sigma^\nu}{2
  \tr(A_\Lambda^\dagger)},\quad
\tilde A_\Lambda=\frac{\tilde\sigma^\mu\Lambda_{\mu\nu}\sigma^\nu}{2
  \tr(\tilde A_\Lambda^\dagger)},\quad
\tilde A_\Lambda^{-1}=\frac{\tilde\sigma^\mu\Lambda_{\nu\mu}\sigma^\nu}{2
  \tr(\tilde A_\Lambda^\dagger)}.
\end{equation}
$A_\Lambda$ and $\tilde A_\Lambda$ can be written in an exponential form
similar to Eq.~(\ref{Lamexp}),
\begin{equation}\label{AtAexp}
A_\Lambda(\omega)=\exp\left(-\frac12
  \omega_{\alpha\beta}a^{\alpha\beta}\right),\qquad
\tilde A_\Lambda(\omega)=\exp\left(-\frac12
  \omega_{\alpha\beta}\tilde a^{\alpha\beta}\right).
\end{equation}
For the exponential coefficients of $A_\Lambda$ one obtains
\begin{equation}
a^{\alpha\beta}=\frac14(e^{\alpha\beta})_{\mu\nu}\sigma^\mu\tilde\sigma^\nu
  =-\frac12(\sigma^\alpha\tilde\sigma^\beta-\sigma^\beta\tilde\sigma^\alpha)
\end{equation}
which can be detailed into
$a^{ij}=\frac i2\epsilon_{ijk}\sigma_k=:-\epsilon_{ijk}b_k$,
$a^{0i}=\frac12\sigma_i=:-e_i$ with
\begin{equation}
b_i=-\frac i2\sigma_i,\qquad e_i=-\frac12\sigma_i,
\end{equation}
where $b_i$ and $e_i$ obey the algebra $\lOR(1,3)$,
\begin{equation}
[b_i,b_j]=\epsilon_{ijk}b_k,\qquad
[b_i,e_j]=\epsilon_{ijk}e_k,\qquad
[e_i,e_j]=-\epsilon_{ijk}b_k.
\end{equation}
For the exponential coefficient of $\tilde A_\Lambda$ one obtains
\begin{equation}
\tilde a^{\alpha\beta}
  =\frac14(e^{\alpha\beta})_{\mu\nu}\tilde\sigma^\mu\sigma^\nu=-\frac12
  (\tilde\sigma^\alpha\sigma^\beta-\tilde\sigma^\beta\sigma^\alpha)
\end{equation}
which gives $\tilde a^{ij}=
-\frac i2\epsilon_{ijk}\sigma_k=:\epsilon_{ijk}\tilde b_k$,
$\tilde a^{0i}=\frac12\tilde\sigma_i=:\tilde e_i$. The generators
\begin{equation}
\tilde b_i=-\frac i2\tilde\sigma_i,\qquad\tilde e_i=\frac12\tilde\sigma_i
\end{equation}
(note the sign changes compared to $b_i,e_i$) obey again the algebra
$\lOR(1,3)$,
\begin{equation}
[\tilde b_i,\tilde b_j]=\epsilon_{ijk}\tilde b_k,\qquad
[\tilde b_i,\tilde e_j]=\epsilon_{ijk}\tilde e_k,\qquad
[\tilde e_i,\tilde e_j]=-\epsilon_{ijk}\tilde b_k.
\end{equation}
Formally, the transitions to the induced Lorentz transformations can be
considered as mappings $\pi:\Lambda\to A_\Lambda$ with
$\pi(e^{\alpha\beta})=a^{\alpha\beta}$ and
$\tilde\pi:\Lambda\to\tilde A_\Lambda$ with
$\tilde\pi(e^{\alpha\beta})=\tilde a^{\alpha\beta}$. Under these mappings,
the generators $J^\eps_i$ and $K^\eps_i$ of $\sol_2^\pm$ are mapped onto the
Chevalley basis. Under $\pi$ one obtains
\begin{eqnarray}
J^\eps_3\to \frac14(1+\eps)\begin{pmatrix}1&0\\ 0&-1\end{pmatrix},&&
K^\eps_3\to \frac14(1-\eps)\begin{pmatrix}1&0\\ 0&-1\end{pmatrix},\nonumber\\
J^\eps_-\to \frac14(1+\eps)\begin{pmatrix}0&0\\ 2&0\end{pmatrix},&&
K^\eps_+\to \frac14(1-\eps)\begin{pmatrix}0&2\\ 0&0\end{pmatrix},
\end{eqnarray}
while under $\tilde\pi$ one obtains
\begin{eqnarray}
J^\eps_3\to-\frac14(1-\eps)\begin{pmatrix}1&0\\ 0&-1\end{pmatrix},&&
K^\eps_3\to-\frac14(1+\eps)\begin{pmatrix}1&0\\ 0&-1\end{pmatrix},\nonumber\\
J^\eps_-\to-\frac14(1-\eps)\begin{pmatrix}0&0\\ 2&0\end{pmatrix},&&
K^\eps_+\to-\frac14(1+\eps)\begin{pmatrix}0&2\\ 0&0\end{pmatrix},
\end{eqnarray}
i.e.\ the same result with $\eps\leftrightarrow-\eps$ and the total sign
interchanged. Again, we are faced with the fact that half of the generators
are mapped to zero. Taking into account the relations to $J^\boxplus_i$ and
$K^\boxplus_i$, one can state that $\pi$ maps to the first component of the
Kronecker sum while $\tilde\pi$ maps to the second component of the Kronecker
sum. Due to Sec.~\ref{sec4}, $\pi$ is the mapping to the left-handed sector,
$\tilde\pi$ the mapping into the right-handed sector.

Using Eqs.~(\ref{ALamdef}) and performing a couple of simple conversions, one
obtains the explicit shape for $A_\Lambda$ for $\Lambda$ of Eq.~(\ref{Lamexpl})
with dependence on $\eps$,
\begin{equation}
A_\Lambda^\eps=
  \begin{pmatrix}e^{(-\eps t+iw)/2}&\frac12(1-\eps)(u-iv)e^{(-\eps t-iw)/2}\\
  -\frac12(1+\eps)(u+iv)e^{(\eps t+iw)/2}&e^{(\eps t-iw)/2}\end{pmatrix}
\end{equation}
($\tilde A_\Lambda^\eps=A_\Lambda^{-\eps}$), which can be rewritten as
\begin{equation}
A_\Lambda^\eps=R_{-\eps t}A_{u,v}^\eps R_{iw},
\end{equation}
where
\begin{equation}
R_x:=\begin{pmatrix}e^{x/2}&0\\ 0&e^{-x/2}\end{pmatrix},\qquad
A_{u,v}^\eps:=\begin{pmatrix}1&\frac12(1-\eps)(u-iv)\\
  -\frac12(1+\eps)(u+iv)&1\end{pmatrix}
\end{equation}
and $\det A_{u,v}^\eps=\det R_x=1$. Using
$(A_\Lambda^{\eps\dagger})^{-1}=(A_\Lambda^\eps)^{-1\dagger}
  =A_\Lambda^{-\eps}=\tilde A_\Lambda^\eps$ and
$(\tilde A_\Lambda^{\eps\dagger})^{-1}=A_\Lambda^\eps$ and
Eq.~(\ref{sigtrans}), one obtains the Lorentz transformation
$\psi_W(\Lambda p)=U(\Lambda)\psi_W(p)$ of the Weyl spinor where
\begin{equation}
U(\Lambda)=\begin{pmatrix}A_\Lambda^\eps&0\\
  0&\tilde A_\Lambda^\eps\end{pmatrix},\qquad
\psi_W(p)=\begin{pmatrix}\psi_-(p)\\ \psi_+(p)\end{pmatrix}.
\end{equation}

\subsection{Representations of the proper orthochronous Lorentz group}
Using the two mappings $\pi:\Lor(1,3)\to\SL(2,\C)$,
$\tilde\pi:\Lor(1,3)\to\SL(2,\C)$ ($\pi:\Lambda\mapsto A_\Lambda$ and
$\tilde\pi:\Lambda\mapsto\tilde A_\Lambda$) and the Kronecker sum, one can
define the representation $(1/2,1/2)$ by
\begin{equation}
\pi^{(1/2,1/2)}(\Lambda):=(\pi\otimes\tilde\pi)(\Lambda\boxplus\Lambda)
  =\pi(\Lambda)\boxplus\tilde\pi(\Lambda),
\end{equation}
for which (and for the choice $\eps=+1$)
\begin{eqnarray}
\pi^{(1/2,1/2)}(E_i)=-\frac12\sigma_i\boxplus\left(+\frac12\sigma_i\right)
  =E^\boxplus_i,&&
\pi^{(1/2,1/2)}(J_i)=\frac12\sigma_i\boxplus\0=J^\boxplus_i,\nonumber\\
\pi^{(1/2,1/2)}(B_i)=-\frac i2\sigma_i\boxplus\left(-\frac i2\sigma_i\right)
  =B^\boxplus_i,&&
\pi^{(1/2,1/2)}(K_i)=\0\boxplus\frac12\sigma_i=K^\boxplus_i.
\end{eqnarray}
Therefore, the map $\pi^{(1/2,1/2)}:\Lor(1,3)\to\SL(2,\C)\otimes\SL(2,\C)$
may replace the similarity transformation via $S$. The benefit of using this
map instead of the similarity transformation is that such a construction can
easily be generalized to a representation $\pi^{(k,l)}$ of the proper
orthochronous Lorentz group.

In proceeding to these general $(k,l)$ representations, the common
eigenvectors $(1,0)^T$ of the set $\{\sigma_3,\sigma_+\}$ and $(0,1)^T$ of the
set $\{\sigma_3,\sigma_-\}$ can be written as states
$|l;m\rangle=|1/2;1/2\rangle$ and $|l;m\rangle=|1/2;-1/2\rangle$,
respectively, with
\begin{equation}
\sigma_3|l;m\rangle=2m|l;m\rangle,\qquad
\sigma_\pm|l;m\rangle=2\rho(l;\pm m)|l;m\pm 1\rangle,
\end{equation}
where $\rho(l;m)=\sqrt{(l-m)(l+m+1)}$. For the general $(k,l)$
representation the states are given by $|k,l;m_k,m_l\rangle$, with
\begin{eqnarray}
\pi^{(k,l)}(J_3)|k,l;m_k,m_l\rangle
  &=&2m_k|k,l;m_k,m_l\rangle,\nonumber\\[3pt]
\pi^{(k,l)}(J_\pm)|k,l;m_k,m_l\rangle
  &=&2\rho(k;\pm m_k)|k,l;m_k\pm 1,m_l\rangle,\nonumber\\[7pt]
\pi^{(k,l)}(K_3)|k,l;m_k,m_l\rangle
  &=&2m_l|k,l;m_k,m_l\rangle,\nonumber\\[3pt]
\pi^{(k,l)}(K_\pm)|k,l;m_k,m_l\rangle
  &=&2\rho(l;\pm m_l)|k,l;m_k,m_l\pm 1\rangle.
\end{eqnarray}
Of these $(2k+1)\times(2l+1)$ states, only those for $\rho(k;-m_k)=0$ or
$\rho(l;m_l)=0$, i.e.\ for $m_k=-k$ or $m_l=l$ are common eigenstates of
$\sol_2^-=\{J_3,J_-\}$ and $\sol_2^+=\{K_3,K_+\}$, respectively, with
$|k,l;-k,l\rangle$ being the common eigenvector for both
algebras~\cite{Saar:2016jbx}.

\subsection{Helicity}
In order to define a helicity
\begin{equation}
H(\vec p)=\hat p\cdot\vec s,\qquad\hat p=\frac{\vec p}{|\vec p\,|},
\end{equation}
one needs a spin vector $\vec s$. This vector can be defined by
$s_i=i\hbar b_i$, because then the commutation relation
$[b_i,b_j]=\epsilon_{ijk}b_k$ for the generators of $A_\Lambda^\eps$ leads
to the usual commutation relation
\begin{equation}
[s_i,s_j]=i\hbar\epsilon_{ijk}s_k
\end{equation}
of an angular momentum algebra. For the three-vector part $\vec p=(0,0,1)^T$
of the momentum vector $p$ generating the Borel subgroup $\Bor(1,3;p)$, one
obtains
\begin{equation}
H(\vec p)=s_3=i\hbar b_3=\frac\hbar2\sigma_3
  =\frac\hbar2\begin{pmatrix}1&0\\ 0&-1\end{pmatrix}.
\end{equation}
Therefore, the common eigenvector $(1,0)^T$ of $\sol_2^+$ has helicity
$h=+\hbar/2$ and the common eigenvector $(0,1)^T$ of $\sol_2^-$ has helicity
$h=-\hbar/2$, in agreement (for $\eps=+1$) with the previous understanding of
left and right.

As $b_i$ is the two-dimensional representation of $B_i$, the concept of
helicity can be generalised to representations $(k,l)$,
\begin{equation}
H^{(k,l)}(\vec p)=\pi^{(k,l)}(i\hbar B_3)=\frac\hbar2
  \pi^{(k,l)}\left(\sigma_3\boxplus\sigma_3\right),
\end{equation}
Applied to the state $|k,l;m_k,m_l\rangle$ one obtains
\begin{equation}
H^{(k,l)}(\vec p)|k,l;m_k,m_l\rangle=\hbar(m_k+m_l)|k,l;m_k,m_l\rangle
\end{equation}
which means that the helicity of this state is $h=\hbar(m_k+m_l)$.

\section{Conclusions and Outlook\label{sec6}}
In this paper, we have calculated the stabiliser group of the proper
orthochronous Lorentz group, which turns out to be the maximal solvable or
Borel subgroup of dimension four. We have explained the continuous transition
between the stabiliser groups of massive and massless particles that describes
the massless limit but fails for exactly massless states. We have dealt with
the system of eigenvectors of the Borel subgroup and shown that the Borel
subgroup can be described by a Kronecker sum of two two-dimensional solvable
groups $\sol_2^\pm$ representing right- and left-handed chirality states.
Finally, in the Chevalley basis we have derived the Weyl and Dirac equations
for massless and massive particles and have defined the helicity of the
massless states. Note that without the generator $T$ such a Kronecker sum of
chiral states would not emerge. The Borel subgroup as the maximal solvable
subgroup of the proper orthochronous Lorentz group provides exactly four
eigenvectors describing these two chiral states, of which the left handed
state is populated by massless fermions, the right handed by antifermions. This
is the physical content of our extension.

Even though the foundations for an explanation of the spin-flip effect are
prepared by this, an exact formulation is not gained here but is aimed for a
future publication. The effect is closely related to the concept of mass which
we want to understand in more detail. In our argumentation we obtained
unexpected help from a not yet published seminal work explaining in detail the
construction of a spin operator by a linear combination of components of the
Pauli--Lubanski pseudovector~\cite{Choi:2018mdd}. Not unexpectedly, the
authors end up with two spin (tensor) operators and corresponding chirality
states that are interchanged under parity transformation. Parity eigenstates
can be constructed as particle or antiparticle compound states. Applying the
Lorentz transformation to the massive states of Ref.~\cite{Choi:2018mdd}, the
parity eigenstates are shown to evolve to solutions of the Dirac equation.

In Ref.~\cite{Choi:2018mdd} it is emphasised that the two spin operators are
neither axial nor Hermitian, and the same holds for the spin operators in the
$(1/2,0)\oplus(0,1/2)$ representation. However, both properties are restored
if applied to particle and antiparticle states. On the other hand, as both
properties are essential for physical states, we can conclude that massless
left- and right-handed states are physical only in the total absence of mass.
This ``gap of (un)physicalness'' as an explanation for the spin-flip effect
has to be investigated in detail.

\subsection*{Acknowledgments}
The research was supported by the Estonian Research Council under Grant
No.~IUT2-27 and by the European Regional Development Fund under Grant No.~TK133.

\end{document}